\documentclass[12pt]{elsarticle}

\usepackage{amssymb}
\usepackage{calrsfs}
\usepackage{ulem}

\journal{Nuclear Physics A}
\begin{document}
\begin{frontmatter}

\title{Nuclear Statistical Equilibrium Equation of State for Core Collapse}

\author[ifin]{Ad. R. Raduta}
\author[lpc]{F. Gulminelli}
\address[ifin]{National Institute for Physics and Nuclear Engineering (IFIN-HH), RO-077125, Bucharest-Magurele, Romania}
\address[lpc]{Universit\'e de Caen Normandie, ENSICAEN, LPC, UMR6534, F-14050 Caen, France}

\begin{abstract}
Extensive calculations of properties of supernova matter are presented, using the extended 
Nuclear Statistical Equilibrium model 
of Ref. \cite{eNSE_PRC2015} based on a statistical distribution of Wigner-Seitz cells modeled 
using realistic nuclear mass and level density tables, complemented 
with a non-relativistic Skyrme functional for unbound particles and beyond drip-line nuclei.
Both thermodynamic quantities and matter composition are examined as a function of baryonic density, 
temperature, and proton fraction, within a large domain adapted for applications in supernova simulations.
The results are also provided in the form of a table, with grid mesh and format compatible with the 
CompOSE platform \cite{compose} for direct use in supernova simulations. 
Detailed comparisons are also presented 
with other existing databases, all based on relativistic mean-field functionals, 
and the differences between the different models are outlined. 
We show that the strongest impact on the predictions is due to the different hypotheses used to 
define the cluster functional and its modifications due to the presence of a nuclear medium.
 
\end{abstract}

\end{frontmatter}

\section{Introduction}

During the in-fall and post bounce stages of the core collapse evolution of massive stars
huge domains of density, temperature and charge fraction are explored.
Matter consists of baryons, leptons (electrons, positrons, neutrinos and antineutrinos) and photons,
and it has a homogeneous/unhomogeneous structure at supra-/sub-saturation densities. 
Leptons and photons interact weakly and are customarily treated as ideal Fermi and, respectively, Bose gases
\cite{Lattimer_NPA_1985}.
Composition and thermodynamics of baryonic matter, generically called equation of state (EoS), is still a matter of research.
The reasons rely on both uncertainties related to the effective interactions and difficulties in the modelling.
The core-collapse supernova dynamics does not only depend on the EoS, but also, and more importantly, 
on the progenitor models, on the  weak interaction rates (electron capture, $\beta$-decay and 
neutrino absorption and scattering), and on the modelling of neutrino transport; 
however all these different aspects are closely inter-correlated, and a reliable modelling 
of matter composition is very important to limit the propagation of uncertainties.

Because of the complexity of core-collapse supernova dynamics \cite{Janka_2012}, 
which requires full relativistic hydrodynamics in three dimensions to be coupled with a complete solution
of the Boltzmann equation for neutrino transport, energetics and composition of nuclear matter 
is implemented via external tables covering within a small mesh the huge ranges of thermodynamic conditions 
explored during the astrophysical evolution. In this way the coupling of very different
length scales is avoided and the sensitivity studies on the different ingredients is much simplified.

The first and so far most intensively used equation of state (EoS) models employed simplifying 
hypothesis on matter composition at densities below the saturation density of symmetric nuclear matter.
Indeed, the Single Nucleus Approximation (SNA) on which Refs. 
\cite{Lattimer_NPA_1985,LS_NPA_1991,HShen_NPA_1998,HShen_ApJSS_2011} rely,
assumes that nuclear matter consists of a homogeneous gas of self-interacting neutrons and protons, 
a free gas of alpha particles, and a unique cluster of nucleons, all of which in thermal, strong, 
but not necessarily weak equilibrium.
It is analogous to the one-component plasma (OCP) description of the catalyzed crust of neutron stars 
\cite{BPS_ApJ_1971}, 
that relies on the energy minimization condition within a solid lattice structure. 
The nuclear cluster treatment, realized either within the Compressible Liquid Drop Model (CLDM) 
\cite{Lattimer_NPA_1985,LS_NPA_1991} or the Thomas Fermi (TF) approximation 
\cite{HShen_NPA_1998,HShen_ApJSS_2011}, allowed to account for in-medium modifications.
Coherent use of the same energy density functional for the unbound nucleon gas and 
the cluster warranted a correct treatment of the transition between unhomogeneous and homogeneous matter, 
taking place at densities slightly lower than the saturation density,
and made possible the first studies on the interplay between star matter EoS and EoS of nuclear matter.
It also allows, at least in principle, to encode in this modelling any recent development of 
effective interactions coming from experimental measurements or ab-initio modelling.

The limit of this approach comes from the fact that most of the time, according to collapse simulations, 
temperatures are above the crystallization temperature and the minimization of the 
thermodynamic potential naturally leads to a whole distribution of different nuclear species. 
The increasing importance of the distribution of nuclear clusters as the temperature increases 
was acknowledged for the first time in Ref. \cite{Lattimer_NPA_1985}, 
where approximate formulas have been proposed.
The limitations related to SNA have been recently addressed in Refs. \cite{eNSE_PRC2015,NSE_SNA}. 
Ref. \cite{NSE_SNA} showed that the cluster distributions can be approximated by the most probable nucleus 
only when they are close to a Gaussian, as it comes out to be the case when one adopts,
for the cluster energy functional, a smooth and continuous function \cite{Botvina_NPA_2010}. 
However, due to the nuclear shell effects, the nuclear free energy is strongly discontinuous up to temperatures 
of the order of a few MeV. 
Under a wide range of thermodynamic conditions, competition between neutron and proton magic numbers leads to 
multi-modal cluster distributions and, thus, a composition that is very different from 
the one predicted by the SNA \cite{NSE_SNA}. 
While this was argued to only marginally affect the EoS \cite{Burrows_ApJ_1984}, important effects 
on the electron capture (EC) rates were reported
\cite{Juodagalvis_NPA_2010,Fischer_EPJA_2014,Magic_PRC_2016,Furusawa_PRC_2017}.
They are obviously due to the nuclear structure dependence of the weak interaction rates.
Considering that EC rates determine the deleptonization dynamics and the global neutrino production 
\cite{Lattimer_PhysRep_2000,Langanke_PRL_2003,Hix_PRL_2003,Janka_PhysRep_2007,Sullivan_ApJ_2016,EC_PRC_2017}, 
it is easy to understand that it is of key importance to account for realistic nuclear distributions.

SNA-related drawbacks can be overcome within the Nuclear Statistical Equilibrium (NSE) approach 
\cite{HNW_AA_1984}, valid at $T \gtrsim 0.5$ MeV.
In the recent years a number of improved EoS models including the full nuclear distribution 
has started to become available 
\cite{Heckel_PRC_2009, RG_PRC_2010, Botvina_NPA_2010, Hempel_NPA_2010, Blinnikov_AA_2011, Furusawa_ApJ_2011,Hempel_ApJ_2012, SHF_ApJ_2013, Furusawa_ApJ_2013, eNSE_PRC2015, Furusawa_NPA_2017,Furusawa_JPG_2017}.
Hybrid EoS models that combine, over complementary density-temperature domains, 
SNA and NSE have been proposed as well
\cite{GShen_PRC_2011,Schneider_PRC_2017} and their EoS tables are publicly available. 
At variance with the original NSE \cite{HNW_AA_1984} model based of the solution of the Saha equations,
the recently proposed extended NSE models effectively account for nuclear interactions among 
the unbound nucleons as well as interactions between the unbound nucleons and nuclei. 
The interaction among the unbound nucleons is accounted for by the employed energy density functional.
The interaction between the unbound nucleons and the nuclei is mimicked via the classical excluded 
volume approximation, 
which prevents different species to occupy the same volume. This excluded volume formalism extends to a 
nuclear distribution the treatment which is employed within the SNA treatment.
The effective nuclear interaction being still largely unknown in dense and strongly isospin asymmetric matter,
the proposed models span different behaviors in both isoscalar and isovector channel.
In addition to this, differences in cluster modelling lead to some model dependence 
\cite{Buyukcizmeci_NPA_2013}, especially
in what regards the chemical composition and the transition to uniform matter.
The impact the nuclear EoS has on collapse evolution can only be assessed by performing simulations 
\cite{Hempel_ApJ_2012,Fischer_EPJA_2014,Furusawa_JPG_2017,Schneider_PRC_2017}. 
To this aim a sufficiently large number of EoS tables have to be available.

To contribute to this collective effort of the nuclear astrophysics community, in this paper we present 
complete EoS tables from the extended NSE model of ref. \cite{eNSE_PRC2015}. 
This table, where the energy functional is taken from Skyrme interactions, can be considered as 
complementary to the existing NSE models which employ relativistic mean field (RMF) parametrizations. 
Indeed it is well known that relativistic (RMF) and non-relativistic (Skyrme) functionals do not cover 
the same range of empirical EoS parameters \cite{Dutra_PRC_2008}. Apart from the choice of the
energy functional, other more technical differences exist among the different models, 
essentially concerning the treatment of clusters which, even if the functional was perfectly known, 
remains a complex many-body problem only solved within strong approximations \cite{Buyukcizmeci_NPA_2013}. 
In the next sections, the different ingredients of the model that can lead to model dependencies 
are explained in detail, and a throughout comparison is presented with other EoS tables available 
in the literature.

As a general result, we observe a good agreement on the different thermodynamic quantities 
despite considerable differences in the composition. We conclude that the EoS uncertainties in 
supernova modelling essentially concern the approximations of the many-body treatment of 
nuclear clusters embedded in a nuclear medium, much more than the present uncertainties in the 
nuclear energy functional.

\section{The model}

The extended NSE model was initially proposed in Ref. \cite{RG_PRC_2010} and subsequently developed in Refs. 
\cite{Raduta_EPJA_2014,eNSE_PRC2015}. Applications on core collapse have been presented in 
Refs. \cite{Magic_PRC_2016,EC_PRC_2017}.
The basic idea of the grand-canonical version of the model~\cite{eNSE_PRC2015}, 
used for the generation of the present tables, is to replace the NSE distribution of non-interacting nuclei 
with a distribution of non-interacting Wigner-Seitz (WS) cells, with appropriate boundary conditions. 
The distribution of clusters is then obtained by factorizing out of the total partition sum 
the free nucleon contribution, which is treated as a self-interacting homogeneous gas 
in the mean-field approximation.

This main simplifying hypothesis of non-interacting WS cells, which is shared with all EoS models 
we are aware of, means that nuclear and, more important, Coulomb interactions among clusters 
are completely disregarded.
The use of Wigner-Seitz cells as degrees of freedom guarantees that the correct zero temperature limit 
is properly recovered~\cite{eNSE_PRC2015}, that is that the predictions of the extended NSE coincide with the 
standard minimization of the energy density in the limit of vanishing temperature. 
Also, this formalism guarantees that the most probable cluster of the distribution exactly coincides with the 
unique cluster of the SNA approximation. Thus, under the thermodynamic conditions where the SNA 
approximation is justified, our approach naturally converges to the SNA thus avoiding possible 
discontinuities when different regimes are matched \cite{Schneider_PRC_2017}. 
Within this statistical treatment, the possible model dependence is uniquely due 
to the expression employed for the free energy of a WS cell, the other equations resulting by 
general statistical mechanics expressions. The expression of the free energy of a WS cell requires a choice 
for the energy functional and contains different approximations, which sometimes involve a certain 
degree of arbitrariness.
These approximations are explained in details in this section. 

The general derivation of the formalism and the main equations can be found in Refs.~\cite{Raduta_EPJA_2014,eNSE_PRC2015}.
Since modifications and improvements were added during the years, in this section we briefly 
recall all the main ingredients which are used in the present version of the code.

\subsection{From distribution of WS cells to distribution of clusters}

The main hypothesis of all extended NSE models including ours is the absence of nuclear and 
Coulomb interactions among the different clusters. In this hypothesis, the system in a given 
thermodynamic condition $(n_B,T,Y_p)$ can be viewed as a collection of non-interacting Wigner-Seitz cells, 
defined as electrically neutral volumes centered on each cluster. 
Each Wigner-Seitz cell $(i)$ contains a single cluster and is characterized by a baryon number $A_i$ 
and atomic number $Z_i$. 
Because of the boundary conditions, the densities of free electrons ($n_e$) and unbound protons ($n_{gp}$) 
and neutrons ($n_{gn}$) in the different cells are the same. 
The cell volume is given by the neutrality condition, $V_i=Z_i/n_e$. 
The total Helmholtz free-energy of the system in a given configuration $k$ is given by:
\begin{equation}
F_{tot}(n_B,T,Y_p)=\sum_i n_i^k F_i,
\end{equation}
where $n_i^k$ is the number of occurrences of the cell $i$ in the total volume for the configuration $k$, 
and $F_i$ is the free energy of the cell.
The grand-canonical partition sum is computed as:
\begin{equation}
Z_{\beta,\mu_n,\mu_p}=\sum_k \exp \left ( -\beta\sum_i n_i^k \left (F_i-\mu_n N_i - \mu_p Z_i \right)\right ),
\end{equation}
where the sum extends to all possible configurations (or microstates, in statistical mechanics language) and
the usual notation $\beta=1/k_B T$ is used. 
The cell free energy $F_i$ depends on the variables of the cell $A_i,Z_i,V_i$ but also on the densities of 
the unbound particles $n_e$, $n_{gn}$,$n_{gp}$ which are common to all cells. 
These quantities are implicitly dependent on the occupations $n_i^k$ through the conservation laws valid for each configuration $k$:
\begin{eqnarray}
n_B&=&\frac{1}{V_{tot}}\sum_i n_i^k A_i ,\\
Y_p n_B&=&\frac{1}{V_{tot}}\sum_i n_i^k Z_i=n_e.
\end{eqnarray} 
For this reason, additional rearrangement terms arise, and the free energy appearing in the probability distributions 
is given by $F_i + \partial F_{i}/\partial n_i|_{n_j}$. 
These rearrangement terms amount to shift the chemical potentials 
with respect to the free system value \cite{eNSE_PRC2015,NSE_SNA}. We account for the shift which is trivially 
due to mass conservation, as it is shown below. 
This same shift appears in the Hempel and Schaffner-Bielich formalism \cite{Hempel_NPA_2010} 
through the excluded volume correction implemented by those authors. 
In principle, extra rearrangement terms should arise from the explicit density dependence of the cluster functional 
due to electron polarization effects. To our knowledge, these effects are neglected in all NSE models, including ours. 

The cell free energy is rearranged such as to sort out the contribution of a uniform gas of unbound particles as:
\begin{equation}
F_i-\mu_n N_i - \mu_p Z_i=F_i^{(e)}-\mu^{(e)}_i + V_i \left (  f_{HM}-\mu_n n_{gn}-\mu_p n_{gp} + f_e-\mu_e n_e\right ). 
\label{gi}
\end{equation}
Here, $f_e(n_e)$ is the free energy density of an ideal gas of electrons, $\mu_e=\partial f_e/ \partial n_e$ and 
$f_{HM}(n_{gn},n_{gp})$ is the free energy density of homogeneous nuclear matter.
Their respective expressions are given by Refs. \cite{Copperstein_NPA_1985,LS_NPA_1991} and
the standard density functional theory \cite{Vautherin_NPA1996}. 
For all the numerical applications presented in this paper and in the associated tables, 
the Skyrme SLy4 functional \cite{SLY4} is used, which provides a good description of binding energies 
and radii of atomic nuclei as well as pure neutron matter as calculated by ab-initio models. 
Equation (\ref{gi}) defines the in-medium modified free energies $F_i^{(e)}$ and chemical potentials 
$\mu^{(e)}_i$ of the clusters, whose expressions are given in the next section. 
The introduction of $G_i^{(e)}=F_i^{(e)}-\mu^{(e)}_i$ allows factorizing the partition sum as:
 \begin{equation}
Z_{\beta,\mu_n,\mu_p}= \left( z_e z_{HM}\right )^{V_{tot}}  Z^{cl}_{\beta,\mu_n,\mu_p}, \label{fact}
\end{equation}
with standard notations for the homogeneous components partition sums, 
$-k_BT\ln z_{HM}=f_{HM}-\mu_n n_{gn}-\mu_p n_{gp}$, $-k_BT\ln z_{e}=f_{e}-\mu_e n_e$.
The cluster partition sum comes out to be identical to the one of the Fisher cluster model \cite{SMM}, 
where however the vacuum expression for the cluster Gibbs energy is replaced by the in-medium modified one 
(here noted by the superscript $(e)$) 
\cite{eNSE_PRC2015},  
\begin{equation}
Z^{cl}_{\beta,\mu_n,\mu_p}=\prod_i \sum_{m=0}^{\infty} \frac{ \left[ \exp \left[-\beta G_i^{(e)} \right]
\right] ^m}{m!}=
\prod_i \exp \omega_{\beta,\mu_n,\mu_p} (i),
\end{equation}
with $ \omega_{\beta,\mu_n,\mu_p} (i)= \exp \left[-\beta G_i^{(e)} \right]$.
The statistical average prediction for the number of occurrences of the cell $i$ is then given by:
\begin{equation}
\langle n_i\rangle = \frac{\partial Z_{\beta, \mu_n, \mu_p}^{cl}}{\partial \left(\beta\mu_i \right)}
=\omega_{\beta,\mu_n,\mu_p} (i).
\end{equation}
A given thermodynamic condition in the grand-canonical ensemble $(\beta,\mu_n,\mu_p)$ 
is associated to well defined values 
of the unbound components densities $n_{q}$ ($q=gn,gp,e$) as $n_q=k_BT\partial \ln z_q/\ln \mu_q$. 
Therefore at a given thermodynamic condition there is a one-to-one correspondence between a WS cell $i$ and 
the cluster species $(A,Z)$ which is present in that cell, implicitly defined by eq.(\ref{gi}). 
This finally gives the cluster distribution as:
\begin{equation}
p_i=\frac{\exp \left(-\beta G_i^{(e)}\right)}{\sum_i \exp \left(-\beta G_i^{(e)} \right)}.\label{pnse}
\end{equation}
The extended NSE numerical code then consists in the self-consistent solution, for a given input set $(T,n_B,Y_p)$, 
of the coupled equations $\langle V\rangle n_B=\sum_i p_i A_i$,  $\langle V\rangle Y_p n_B=\sum_i p_i Z_i$, 
where $p_i$ is given by eq. (\ref{pnse}).

\subsection{The WS free energy and the definition of e-clusters}
\label{ssec:WScellandcl}

Let us consider a WS cell of volume $V$ composed of a dense component (or "cluster") 
of atomic and mass number $A_r,Z_r$, and a uniform density of unbound protons $n_{gp}$, neutrons $n_{gn}$ 
and electrons $n_e$. The free energy of the cell is given by 

\begin{eqnarray}
F&=& E^{vac}(A_r,Z_r)-TS^{vac}(A_r,Z_r) \\ \label{fcell}
&+&(V-V_0)f_{HM}(n_{gp},n_{gn}) +\delta E_{surf} +\delta E_{Coul} + V f_e(n_e), \nonumber
\end{eqnarray}
where $E^{vac}(S^{vac})$ is the vacuum energy (entropy) of the cluster, $V_0$ is the cluster volume related 
to the average cluster density $n_0$ by $V_0=A_r/n_0$, 
$\delta E_{Coul}$ is the electron-electron and electron-cluster Coulomb interaction energy and $\delta E_{surf}$ 
is the modification of the cluster surface energy due to the interaction with the external gas.  
The reduced volume $\left( V-V_0 \right)$ 
available to the unbound component accounts for the excluded volume effect \cite{Hempel_NPA_2010} 
and is sometimes referred to as "coexisting phase approximation" \cite{Avancini_PRC_2008,Avancini_PRC_2009}.

To achieve the decomposition of eq.(\ref{gi}) we write:
\begin{eqnarray}
F&=&E^{(e)}- T S^{(e)}+ V \left ( f_{HM} + f_e \right ), \label{fi} \\
A&=&A_e +(n_{gp}+n_{gn}) V \\
Z&=&Z_e +n_{gp} V
\end{eqnarray}
where $A_e (Z_e)$ represent the number of bound nucleons (protons), and  $F^{(e)}=E^{(e)}- T S^{(e)}$ 
gives the cluster free energy, modified by the interactions with the unbound nucleons and electrons.

The cluster chemical potential appearing in  eq.(\ref{gi}) is thus given by
\begin{equation}
\mu_i=\mu_p Z_{i,e} +\mu_n (A_{i,e}-Z_{i,e}) 
\end{equation}

We can see from eq.(\ref{pnse}) that the equilibrium abundances do not depend on the baryonic and atomic number of 
the dense component $A_r,Z_r$, but only on its bound part, or  "e-cluster"~\cite{Panagiota_PRC2013},
given by the left over part of the cluster after subtracting the contribution
of the nucleons of the gas they are embedded in~\cite{Panagiota_PRC2013},
\begin{eqnarray}
A_e=A_r \left(1-\frac{n_g}{n_0} \right), Z_e=Z_r \left(1-\frac{n_{gp}}{n_{0p}} \right),
\label{eq:AeZe}
\end{eqnarray}
where $n_0$ ($n_{0p}$) is the bulk total (proton) density of the cluster.
Comparing eq.(\ref{gi}) and eq.(\ref{fi}) we get for the in-medium modified energy:
\begin{eqnarray}
E^e=E^{vac}+\delta E_{bulk}+\delta E_{Coul}+\delta E_{surf},
\label{eq:Ee}
\end{eqnarray}
where 
$\delta E_{bulk}=-\epsilon_{HM} \frac{A_r}{n_0}$ and $\epsilon_{HM}(n_{gn},n_{gp})$ is the energy density of the nucleons in the gas.

We can see that a bulk nuclear energy shift naturally arises from the factorization condition 
of the partition sum eq. (\ref{fact}). 
This shift is due to the excluded volume appearing in eq. (\ref{fcell}), but it can also be physically interpreted as a 
Pauli blocking shift in the Thomas-Fermi approximation \cite{Roepke_PRC_2015,Pais_PRC_2018}: 
if we consider that the continuum single particle states  of the cluster can be approximated  by plane waves, 
such states are occupied by the unbound component and must therefore be excluded in the energy evaluation of the cluster. 
This approximation to the Pauli-blocking energy shift is only justified for heavy clusters, while the shifts should be 
calculated microscopically in the case of light particles \cite{Roepke_PRC_2015}.
Such microscopic shifts are included for deuterons, tritons, helions and $\alpha$ particles in the NSE model by 
Furusawa et al. \cite{Furusawa_ApJ_2011,Furusawa_ApJ_2013, Furusawa_NPA_2017,Furusawa_JPG_2017}. 
A phenomenological expression inspired by the microscopic shifts \cite{Typel_PRC_2010} 
is used in the gRDF model of Ref. \cite{Pais-Typel}
for all clusters, instead than the excluded volume effect. 
In this work, we do not attempt to make the distinction between light and heavy clusters, and use the simple excluded-volume shift 
for all clusters, similar to Ref. \cite{Hempel_NPA_2010}. For a comparison between the excluded volume mechanism 
and the gRDF prescription, see Ref. \cite{Pais-Typel}.

The extra in medium surface correction $\delta E_{surf}$ reflects a possible modification of the cluster surface tension 
due to the presence of an external nucleon gas. Several authors effectively include this effect in the isospin dependence 
of the surface tension \cite{LS_NPA_1991,Steiner_2005,Newton_ApJSS_2013}.
However more complex dependencies on both cluster size and composition and gas are observed in self-consistent Thomas-Fermi 
calculations in beta-equilibrium matter \cite{Centelles_NPA_1998,Douchin_NPA_2000,Sil_PRC2002}. 
Inspired by these self-consistent calculations, simple analytic expressions are proposed of an in-medium modification 
of the surface tension \cite{Pais_PRC_2016,Grams_PRC_2017}.
In the context of NSE models, an explicit correction depending both on the gas density and on the temperature is introduced 
in the NSE model by Furusawa et al. \cite{Furusawa_ApJ_2011,Furusawa_ApJ_2013, Furusawa_NPA_2017,Furusawa_JPG_2017}.
Thorough investigation of this aspect is under work and will be published elsewhere. 
As such, the approximation $\delta E_{surf} \approx 0$ will be done through this paper.  

The Coulomb energy shift due to electron screening is treated in our model, as frequently done in the literature,
within the Wigner-Seitz approximation,
\begin{equation}
\delta E_{Coulomb}=a_c f_{WS} Z_r^2/A_r^{1/3},
\label{eq:ECoulomb}
\end{equation}
with
\begin{equation}
f_{WS}=\frac32 \left[\frac{n_{e}}{n_{0p}} \right]^{1/3}
-\frac12 \left[ \frac{n_{e}}{n_{0p}}\right],
\end{equation}
and the Coulomb parameter $a_c=0.69$.
 
For the energy functional of the clusters in vacuum, $E^{vac}$, the following choices are made.
For nuclei for which experimental masses are known, the AME2012 mass tables of Audi {\it el al.} 
\cite{Audi_2013} are used. 
Then, up to the drip lines, evaluated masses of the 10-parameter model by Duflo and Zuker \cite{DZ10}, 
here after referred to as DZ10, are employed. 
Beyond drip lines, nuclear binding energies are described according to the Liquid Drop Model (LDM)-
like parametrization of Ref. \cite{Danielewicz}, 
which accounts for isospin effects in the surface tension including the effect of 
extra neutrons in the neutron skin \cite{Steiner_2005}. This expression is modified in two respects.
First, a phenomenological pairing term, $\Delta(A)=\pm 12/\sqrt{A}$, where $+(-)$ corresponds to
even-even (odd-odd) nuclei, is added.
Then, two correction terms are included such as to smoothly match, for each isotopic chain, 
the liquid-drop predictions with the limiting values of DZ10.
Based on Hartree-Fock calculations liquid-drop parameters are proposed in Ref. \cite{Danielewicz} 
for several dozens of Skyrme effective interactions.
For the sake of consistency of the energy functional, we use the parametrization corresponding to the 
same effective interaction as the one employed for describing the homogeneous gas, namely SLy4 \cite{SLY4}.
Inclusion of nuclei beyond drip-lines is motivated by the fact that, in medium, other species than those
existing in vacuum may exist \cite{Sil_PRC2002} and accounting for them might, in principle, modify 
the sharing of matter between clusterized and homogeneous components as well as isotopic abundances.
The allowed mass range of clusters is $2 \leq A \leq 300$. 
Note that, at low temperatures and densities close to the transition to homogeneous matter,
larger structures can be formed if the allowed mass range is extended accordingly \cite{Furusawa_ApJ_2011}.
Their existence is nevertheless much dependent on the in-medium surface modification of the energy functional 
and additional shape degrees of freedom, all of which poorly known. 

The  bulk total $n_0$ (proton $n_{0p}$) cluster  density is taken to be the total (proton) number density 
of saturated nuclear matter of isospin asymmetry $\delta=\left(1-2 n_{0p}/n_0 \right)$ . 
It can be expressed ~\cite{Panagiota_PRC2013}    
as a function of the saturation density of symmetric matter $n_0(0)$  by the equation,
\begin{equation}
n_0(\delta)=n_0(0) \left(1-\frac{3 L_{\rm{sym}} \delta^2}{K_{\rm{sat}}+K_{\rm{sym}} \delta^2} \right),
\label{eq:rho0_delta}
\end{equation}
where $L_{\rm{sym}}$, $K_{\rm{sym}}$ and $K_{\rm{sat}}$
represent the slope and curvature of the symmetry energy $J_{\rm{sym}}$ and, respectively, 
the incompressibility of nuclear matter, all calculated for symmetric saturated matter. 

Due to skin effects, the bulk isospin asymmetry $\delta$ obviously differs from the total isopin asymmetry, 
$(1-2Z_r/A_r)$. For the case of a nucleus in the vacuum (where $Z_r=Z_e$, $A_r=A_e$), 
the following analytic expression has been derived within the liquid drop model~\cite{LD_NPA1980},
\begin{equation}
\delta=\delta_0=\frac{1-\frac{2Z_e}{A_e}+\frac{3a_c}{8Q} \frac{Z_e^2}{A_e^{5/3}}}{1+\frac{9 J_{\rm{sym}}}{4Q} \frac{1}{A_e^{1/3}}},
\label{eq:deltabulk_vacuum}
\end{equation}
where contributions of the symmetry energy, surface stiffness and Coulomb are readily identified 
in addition to the size dependence.
$Q$ represents the surface stiffness coefficient.
For the general case of a nucleus in a dilute medium, of interest here,
we employ the expression proposed in Ref.~\cite{Panagiota_PRC2013},
\begin{equation}
\delta(n_g,n_{gp})=\left[1-\frac{n_g}{n_0(\delta)} \right]\delta_0(Z_e,A_e)+
\frac{n_g}{n_0(\delta)}\left( 1-\frac{2n_{gp}}{n_g}\right).
\label{eq:deltabulk_medium}
\end{equation}

The entropy term includes both translational and internal degrees of freedom,
\begin{equation}
S^{(e)} \left(A_r,Z_r,n_g, n_{gp} \right)= \ln V + \ln c_{T}(A_r,Z_r)
+\frac32  \ln A_e,
\end{equation}
where $c_{T}=g_T(A_r,Z_r)(mT/(2\pi\hbar^2))^{3/2}$, 
$m$ denotes the mass of a nucleon and the internal state partition sum is:
\begin{equation}
g_T(A_r,Z_r)=\sum_{i, discrete} \left(2 J_i+1 \right) \exp \left(-E^*/T \right)+
\int_{(cont)} dE^* \rho_{A_r,Z_r}(E^*) \exp \left ( -E^* /T \right ). 
\label{eq:degen}
\end{equation}

Different corrections must be applied to the cluster entropy in order to avoid double counting 
with the gas states \cite{few,tubbs,rauscher,Bonche_NPA_1984,Bonche_NPA_1985}. 
First, the translation term must  only be computed for the bound part
$A_e$  of the cluster. Moreover, the gas states should be subtracted from the internal state partition sum. 
The simplest approach \cite{few}  consists in cutting the partition sum at the nucleon separation energy, 
such as to exclude all continuum states.  This is a crude treatment, 
while a thermodynamically consistent subtraction of the gas partition sum 
was proposed in Ref. \cite{Bonche_NPA_1984,Bonche_NPA_1985,tubbs,rauscher}. 
If the interaction in the gas is neglected, the subtraction can be done analytically, and it was shown in Refs. 
\cite{tubbs,rauscher} that the simple approach of Ref. \cite{few} underestimates the partition sum, 
which can be easily understood  because the resonant states are also cut in that procedure. 
However, an exact state counting in the self-consistent mean-field approximation 
in the case of nuclei beyond the dripline at zero temperature was presented in Ref. \cite{Panagiota_PRC2013}, 
showing that the correct number of particles is obtained for the cluster if all the continuum states are excluded, 
and the contribution of resonant states is relatively small. 
Because of the complexity of the issue, in this work we stick to the simplest procedure 
of cutting the internal energy partition sum at the  minimum between the
average neutron and proton separation energies,  
$\langle S \rangle =\min \left( \langle S_n \rangle, \langle S_p \rangle\right)$.

The full list of low-lying resonances for nuclei with $4 \leq A \leq 10 $ has been considered.
For the level density $\rho(A_r,Z_r)(E^*)$ we used the realistic expression of Ref. \cite{Bucurescu2005}, 
fitted on experimental data.
For more details see Ref. \cite{eNSE_PRC2015}.

\subsection{Thermodynamic quantities}

The total and proton number densities are given by:
\begin{eqnarray}
n_B&=&n_{g}+\frac1V \sum_{A_r,Z_r} A_e n(A_r,Z_r)\\ 
Y_p n_B&=&n_{gp}+\frac1V \sum_{A_r,Z_r} Z_e n(A_r,Z_r).
\label{eq:conserv}
\end{eqnarray}

Clusterized phase pressure, entropy density and internal energy density may be readily calculated from their
thermodynamic definitions. The following expressions are obtained for the pressure, 
\begin{equation}
p_{cl}=T \frac{\partial \ln Z_{\beta,\mu_n,\mu_p}^{cl}}{\partial V}=\frac TV \ln Z_{\beta,\mu_n,\mu_p}^{cl}=
\frac TV \sum_{A_r,Z_r} n(A_r,Z_r),
\label{eq:pcl}
\end{equation}
entropy density,
\begin{eqnarray}
s_{cl}&=&\frac 1V \frac{\partial(T \ln Z_{\beta,\mu_n,\mu_p}^{cl})}{\partial T} \nonumber \\
&=&
\frac 1V \sum_{A_r,Z_r} n(A_r,Z_r) \left( \frac52 +T \frac{\partial \ln g_T(A_r,Z_r)}{\partial T}
+\frac{E^e(A_r,Z_r)-\mu_n (A_e-Z_e)- \mu_p Z_e}{T}
\right),
\label{eq:scl}
\end{eqnarray}
and internal energy density,
\begin{equation}
e_{cl}=\frac 1V \sum_{A_r,Z_r} n(A_r,Z_r) \left(\frac32 T+\langle E^*(A_r,Z_r)\rangle+ E^e (A_r,Z_r)
 \right),
\label{eq:ecl}
\end{equation}
where the average excitation energy of the cluster $(A,Z)$ is,
\begin{equation}
\langle E^*(A_r,Z_r)\rangle=\frac{\int_{0}^{S(A_r,Z_r)} dE^* E^* \rho_{A_r,Z_r}(E^*) \exp(-E^*/T)}{\int_{0}^{S(A_r,Z_r)} dE^* \rho_{A_r,Z_r}(E^*) \exp(-E^*/T)} 
=T^2 \frac{\partial \ln g_T \left(A_r,Z_r \right)}{\partial T}.
\end{equation}

Baryonic pressure, entropy and energy densities are obtained by summing up clusterized and 
homogeneous phases contributions,
\begin{eqnarray}
p_B&=&p_{cl}+p_g \nonumber \\ 
s_B&=&s_{cl}+s_g \nonumber \\
e_B&=&s_{cl}+e_g.
\end{eqnarray}
Note that free volume corrections do not appear explicitly, as in Ref. \cite{Hempel_NPA_2010},
as they have been already taken into account in the definition of e-clusters.

Total pressure, entropy and energy densities are obtained by adding to the baryonic quantities the 
contributions of the electron (including contribution of positrons) and photon gases.
Analytic expressions for the electron gas with temperature larger than 1 MeV and the photon gas are given in Ref. \cite{LS_NPA_1991}.
Expressions for the electron gas at $T \lesssim 1$ MeV have been proposed in Ref. \cite{Copperstein_NPA_1985}.
The total values of the fundamental thermodynamic quantities thus write,
\begin{eqnarray}
p_T&=&p_B+p_{el}+p_{\gamma}-\sum_{A,Z} n(A,Z) a_c \frac{Z^2}{2 A^{1/3}} 
\left[ \frac{n_{el} A}{n_0(\delta) Z}- \left( \frac{n_{el} A}{n_0(\delta) Z}\right)^3\right], \nonumber \\ 
s_T&=&s_B+s_{el}+s_{\gamma}, \nonumber \\
e_T&=&e_B+e_{el}+e_{\gamma}.
\end{eqnarray}
We note the extra negative pressure term coming from the Coulomb lattice and
remind that lattice contribution to energy and entropy enters in eqs. (\ref{eq:ecl}) and (\ref{eq:scl})
via the Coulomb energy shift, eq. (\ref{eq:ECoulomb}).

\subsection{Transition to homogeneous matter}

At densities of the order of $n_0/2-2 n_0/3$ the nonuniform nuclear matter phase is replaced
by a uniform phase which persists up to several times the value of symmetric saturated nuclear matter density.
Physically this transition occurs in order to minimize the system free energy.
The exact value of the transition density and the way in which it takes place, {\em i.e.}
via phase coexistence or not, depend on a number of issues as 
effective interactions, shape degrees of freedom and in-medium surface modification of the energy functional,
which make the phenomenology strongly model dependent.

For the purpose of building an EoS database suitable for astrophysics use, the details of the transition
are less important than the thermodynamic stability and consistency.
As such, for fixed values of $Y_p$ and $T$ the clusterized phase is computed, as described in the above sections,
up to maximum density where the NSE procedure still converges, typically $4 \cdot 10^{-2} - 9 \cdot 10^{-2}$ fm$^{-3}$.
Homogeneous matter is supposed to onset, independently on temperature and proton fraction, at $n_t=10^{-1}$ fm$^{-3}$.
For intermediate values of density, chemical composition and thermodynamic observables are computed 
by linear interpolation between the boundary values. 

\section{Results}
\label{section:results}

EoS databases are usually delivered as three dimensional tables of baryon number density, charge fraction
and temperature. In order to emphasize the transition from clusterized matter to homogeneous matter or
in medium cluster dissolution matter, composition and thermodynamic quantities are 
most frequently plotted and discussed as a function of baryon number density 
at fixed values of the charge fraction and temperature 
\cite{Hempel_NPA_2010,Furusawa_ApJ_2011,GShen_PRC_2011,Furusawa_ApJ_2013,Furusawa_NPA_2017,Furusawa_JPG_2017,Schneider_PRC_2017}.
Evolution as a function of temperature when the values of baryon number density and charge fraction are fixed
is considered mainly when the focus is put on the effects of in-medium interactions,
as is the case of cluster mass fractions from Ref. \cite{Buyukcizmeci_NPA_2013,Pais-Typel}.
As the whole body of literature shows, in the sub-saturation domain and for moderate values 
of the charge fraction, thermodynamic quantities bear little sensitivity to the employed effective interactions, 
nuclear cluster definition,
approximations (NSE vs SNA) and degree of sophistication of the approaches.
This is due to the fact that the thermodynamic variables of a clusterised medium are 
largely determined by the properties of nuclei in vacuum, which are relatively well known and
upon which all effective interactions have been fitted.
The situation is different for matter composition. 
Effects of the cluster definition in terms of maximum size, isospin asymmetry and excitation energy,
density dependence of the symmetry energy,
in-medium interactions of light nuclei \cite{Pais-Typel}, 
evolution of shell effects far from stability \cite{Magic_PRC_2016} 
or temperature \cite{Furusawa_NPA_2017} etc. have been identified. 
Though systematic analyses are not available yet, these effects are
expected to influence the core collapse evolution via the neutrino opacity, electron capture rates and 
energy dissipation of the shock wave.

Two representations are used in this work for investigating the results of our NSE model.
For general survey of global chemical and energetic behaviors (subsections \ref{ssec:compo} and \ref{ssec:thermo}), 
we prefer plots as a function of charge fraction 
for constant values of the two remaining grid parameters, $T$ and $n_B$.
This view offers technical advantages. First, it allows to straightforwardly see the effect of $Y_p$.
Then, for sufficiently different values of $T$ and $n_B$, the curves are outdistanced, 
which enhances the plots readability. 
When confronting our results with those of other models in the literature (subsection \ref{ssec:comparison}) 
we prefer the investigation as a function of baryonic number density when $T$ and $Y_p$ are fixed.
Different thermodynamic conditions are considered:
in subsections \ref{ssec:compo} and \ref{ssec:thermo} we focus on $T$ and $n_B$-values under which 
the employed microphysics hypothesis and approximations are well justified.
In subsection \ref{ssec:comparison} we focus on states populated in proto-neutron stars and late stage
evolution of core-collapse, where the differences in microphysics treatments can be maximized.

\subsection{Composition}
\label{ssec:compo}

\begin{figure}
\includegraphics[angle=0, width=0.99\textwidth]{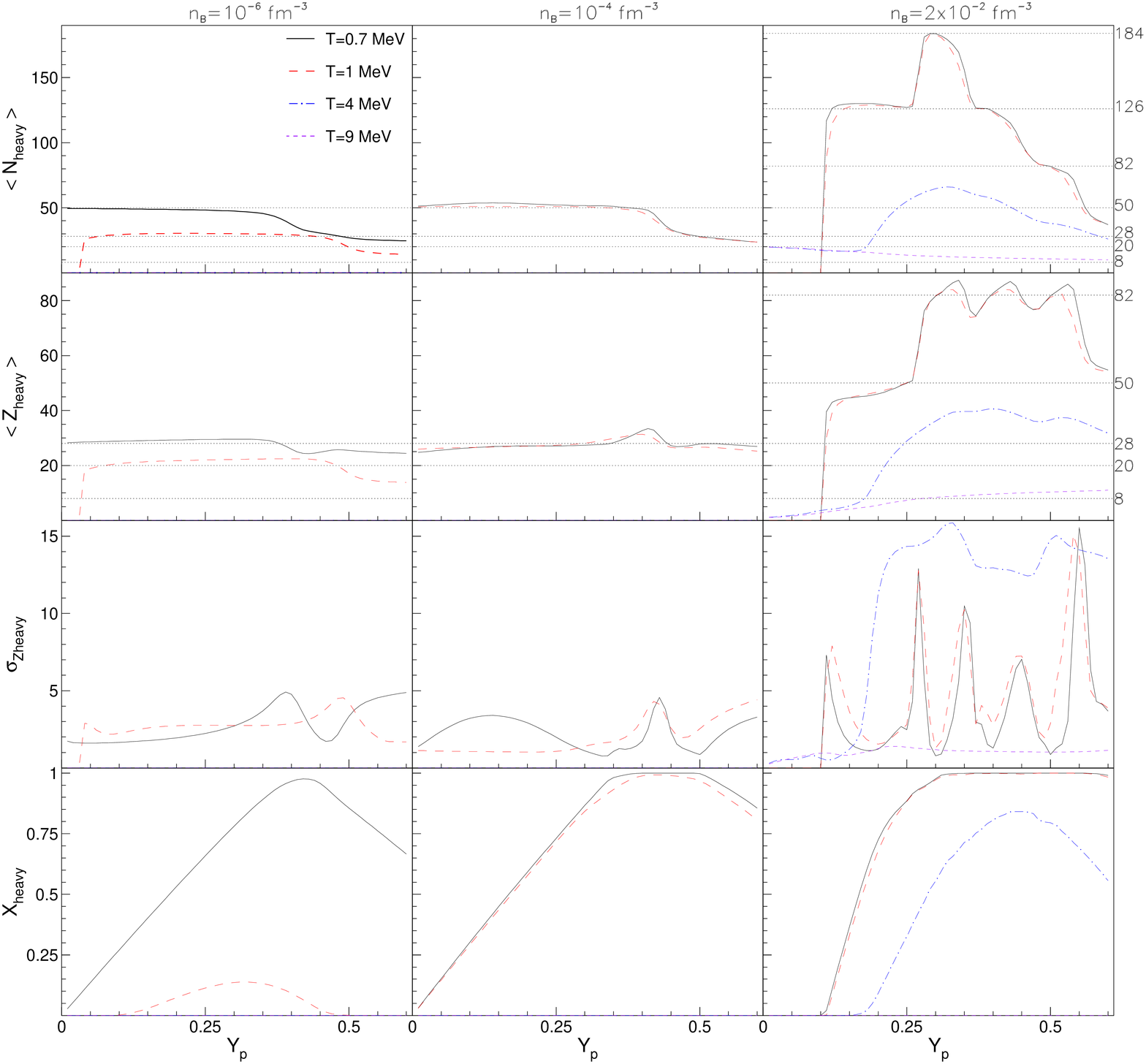}
\caption{Average neutron (1st row panels) and proton (2nd row panels) numbers of nuclei with $A \geq 20$
as a function of proton fraction for fixed values 
of the baryon number density and temperature, see legend.
The 3rd and 4th row panels give, under the same conditions, 
the standard deviation of the proton number, $\sigma_{Z_{heavy}}$, and, 
respectively, the fraction of mass bound in these nuclei.
$N$ and $Z$ magic numbers are marked on the 1st and, respectively, 2nd row panels by horizontal dotted lines.
The values are given at the r.h.s. of the right Y-axis.
}
\label{fig:NhZh}
\end{figure}

\begin{figure}
\includegraphics[angle=0, width=0.99\textwidth]{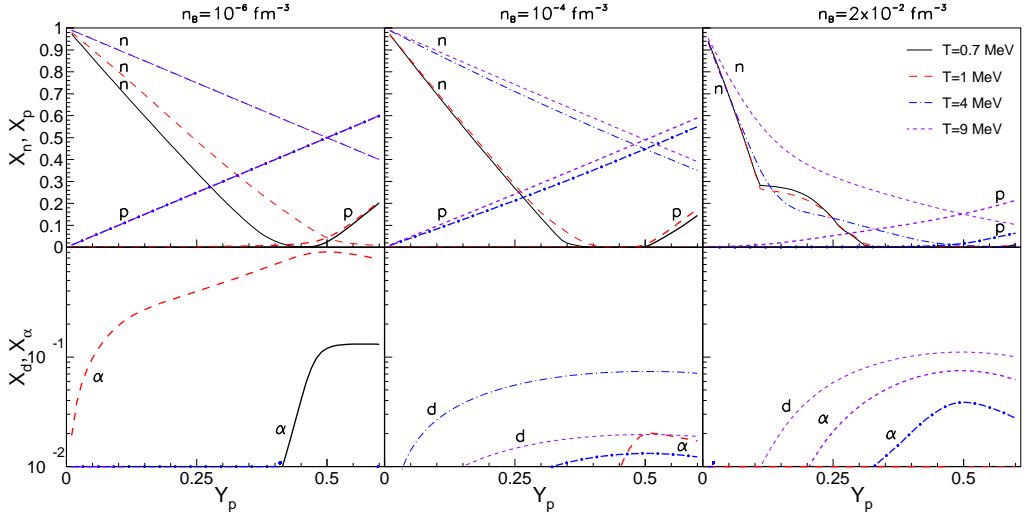}
\caption{Upper (lower) panels: neutron (deuteron) (thin lines) 
and proton ($\alpha$-particle) (thick lines) mass fraction 
as a function of proton fraction for fixed values 
of the baryon number density and temperature, see legend.}
\label{fig:Xnuc}
\end{figure}

Figs. \ref{fig:NhZh} and \ref{fig:Xnuc} illustrate matter composition as a function of $Y_p$
for three representative values of the baryon number density, $n_B=10^{-6}$, $10^{-4}$ and $2 \cdot 10^{-2}$ fm$^{-3}$
and four values of the temperature, $T=0.7$, 1, 4 and 9 MeV.

Fig. \ref{fig:NhZh} depicts the average neutron and proton numbers of clusters 
of mass number $A \geq 20$ (dubbed as "heavy clusters" in the following) together with
the standard deviation of the proton number, $\sigma_{Z_{heavy}}$, and fraction of mass bound
in these nuclei, $X_{heavy} = \sum_{A \geq20,Z} A n(A,Z)/n_B$.
The standard deviation of the neutron number, $\sigma_{N_{heavy}}$, (not plotted)  shows features similar
to those of $\sigma_{Z_{heavy}}$.

As one may notice, for all thermodynamic conditions under which they are produced,
heavy cluster population is determined by the competition between neutron and proton magic numbers.
At low densities and temperatures (e.g. $n_B=10^{-6}, 10^{-4}$ fm$^{-3}$ and $T$=0.7, 1 MeV)
most abundant nuclei have neutron and proton numbers close to $N=8, 20, 28, 50$ and, respectively,
$Z=8, 20, 28$.
The relatively small values of $\sigma_{Z_{heavy}} \lesssim 5$ are explained by the small difference between
the two most frequently competing $Z$-magic numbers, 20 and 28, and the dominance of one of the peaks.
Given the increased variety of competing $N$-magic numbers, slightly larger values characterize
$\sigma_{N_{heavy}}$.
As the density increases, much more numerous and massive clusters are populated up to higher
temperatures. Depending on thermodynamic conditions, including the proton fraction, 
the abundance peaks are the result of competition between a broad range of $N$ and $Z$ 
magic numbers: $N$ =8, 20, 28, 50, 82, 126, 184 and $Z$=8, 20, 28, 50, 82, 114 
(the last superheavy shell closure $N=184,Z=114$ obviously depends on the mass model employed, 
here DZ10 \cite{DZ10}). 
This feature explains the large values of both $\sigma_{Z_{heavy}}$ and $\sigma_{N_{heavy}}$ as well as their
steep evolution with $Y_p$.
Nuclide abundances dominated by magic numbers and, consequently,
large values of the standard deviation of heavy cluster mass distributions 
have been already signaled in Ref. \cite{Furusawa_ApJ_2011}.  

Whether heavy clusters bind a significant amount of matter or not depends on thermodynamic conditions 
(bottom panel of Fig. \ref{fig:NhZh}): 
around isospin symmetry and for the lowest values of temperature $X_{heavy}$ exhausts a large fraction of matter
even at densities as low as $10^{-6}$ fm$^{-3}$.
On the contrary, at high temperatures or in very neutron-rich matter practically no heavy cluster exists.

The mass fractions of neutrons, protons and the two most abundant light clusters, $d$ and $^4$He,
are plotted in Fig. \ref{fig:Xnuc}. 
The top panels of this figure give mass fractions of unbound neutrons and protons.
Both $X_p$ and $X_n$ show strong sensitivities to $T$ and $n_B$.
At high temperatures, unbound nucleons exist over large domains of density and proton fraction
and their abundances monotonically increase (decrease) with $Y_p$.
At variance, at low $T$-values unbound neutrons (protons) exist only in neutron (proton)-rich matter.
At the highest density, the change of slope of the neutron fraction is due to the abrupt appearance 
of the heavy clusters (see Fig.\ref{fig:NhZh} above), which are close to isospin symmetry.
$d$ and $^4$He abundances,  represented in the bottom panels of Fig. \ref{fig:Xnuc},
present a complex and non-trivial evolution with temperature, baryonic number density and proton fraction.
For the lowest considered density, $n_B=10^{-6}$ fm$^{-3}$, $^4$He is produced only at the lowest temperatures.
For $T= 0.7$ MeV, $^4$He exists only in isospin symmetric matter and the associated mass fraction 
barely exceeds 13\%.
The $T= 1$ MeV results show that, quite interesting, $^4$He exists even in very asymmetric matter.
Indeed, at $Y_p \approx 0.1$, $X_{\alpha} \approx 10\%$ while values as high as $X_{\alpha} \approx 80\%$
are attained for $Y_p \approx 0.5$.
For the other two values of baryonic density, significant  $\alpha$-production occurs at higher temperatures, 
e.g $T$=9 MeV for $n_B=2 \cdot 10^{-2}$ fm$^{-3}$,
extends over limited domains of $Y_p$ and gets maximized in symmetric matter.
$d$-production shows features qualitatively similar to those of $\alpha$-production.
Quantitatively, higher densities and temperatures are required to produce $d$ than $\alpha$. 
One may notice that, under specific thermodynamic conditions, the loosely bound $d$ 
may dominate over the strongly bound $^4$He, as already 
observed in previous works \cite{Pais_PRC_2018,Typel_PRC_2010,Avancini2012,Sedrakian_2017}. 
This is the case of $T=4$ and 9 MeV and $n_B=10^{-4}$ fm$^{-3}$ and
$T=9$ MeV and $n_B=2 \cdot 10^{-2}$ fm$^{-3}$, irrespective the value of $Y_p$.
As for $\alpha$, for all considered $n_B$ and $T$-values, the mass fraction of the isospin symmetric deuteron
gets maximized in symmetric matter. 

\begin{figure}
\includegraphics[angle=0, width=0.99\textwidth]{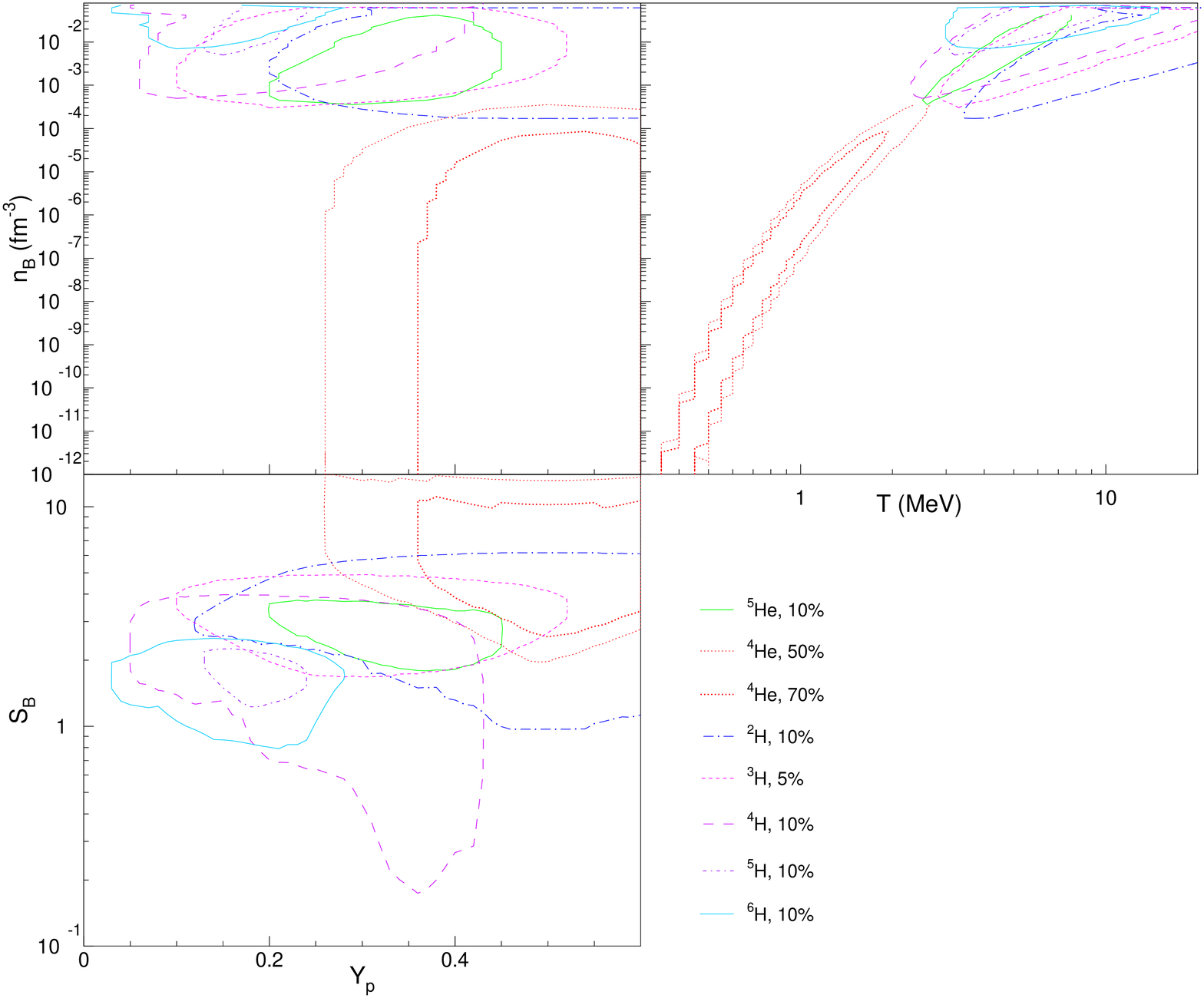}
\caption{Domains in $(n_B,Y_p)$ (top left), $(n_B,T)$ (top right)
  and $(S_B,Y_p)$ (bottom left)
  where the mass fraction of $^2$H, $^3$H, $^4$H, $^5$H, $^6$H and
  $^4$He and $^5$He exceeds certain values, see legend.}
\label{fig:HeandHe}
\end{figure}

Population of $d$ and $^4$He is further considered in Fig. \ref{fig:HeandHe} 
together with that of other isotopes of H and He in terms of mass fraction 
as a function of $T$, $n_B$, $Y_p$ and baryonic entropy per baryon $S_B$. 
The different contours correspond to the mass fraction thresholds shown in the legend. 
We note that by far the most abundant light cluster is $^4$He. 
Its mass fraction exceeds 70\% over considerable domains of 
baryonic density ($10^{-12} \lesssim n_B \lesssim 10^{-4}$ fm$^{-3}$) and 
proton fraction ($0.35 \lesssim Y_p \lesssim 0.6$)
for temperatures up to a couple MeV 
($0.3 \lesssim T \lesssim 2$ MeV), which correspond to entropies per baryon ranging from 3 to 10.
Mass fractions larger than 10\% are obtained for $^5$He, $^4$H, $^5$H, $^6$H
for $T \gtrsim 2$ MeV, $n_B \gtrsim 10^{-4}$ fm$^{-3}$ over various domains of $Y_p$,
which correspond to few $S_B$. 
As easy to anticipate, extreme neutron-rich nuclei ({\it e.g.} $^5H$, $^6H$)
are preferentially produced in neutron-rich environments ($Y_p \lesssim 0.3$).
Indeed, as already observed in Ref. \cite{eNSE_PRC2015,Burrello}, in very asymmetric matter 
and beta-equilibrium matter at high density and temperature, heavy hydrogen and helium isotopes 
strongly dominate over the isospin symmetric $d$ and $^4$He.

\subsection{Thermodynamic quantities}
\label{ssec:thermo}

\begin{figure}
\includegraphics[angle=0, width=0.99\textwidth]{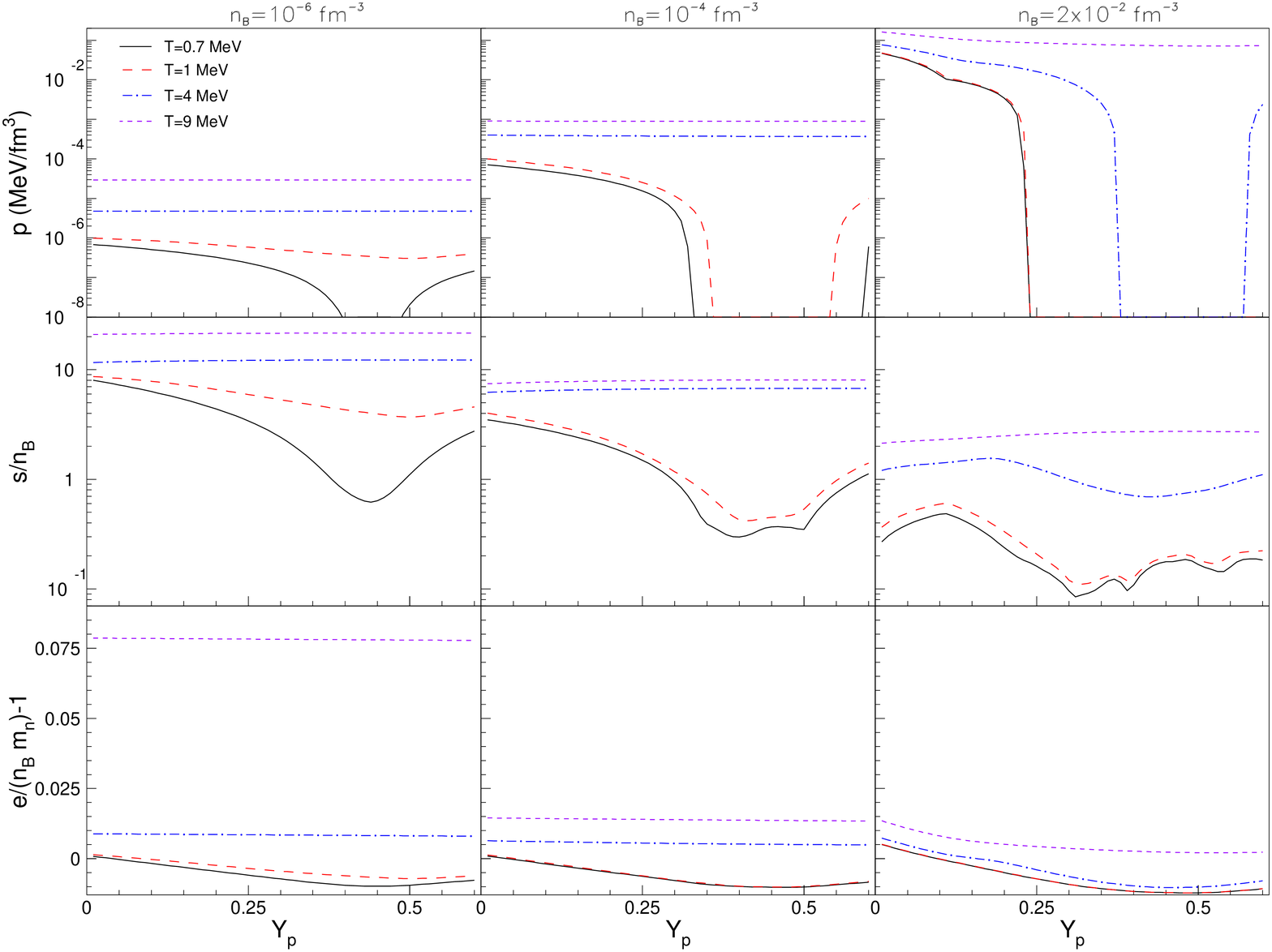}
\caption{Baryonic+lattice pressure (top), entropy per baryon (middle) and shifted
energy per baryon (bottom) as a function of proton fraction for fixed values of
the baryonic number density and temperature, see legend.}
\label{fig:baryon_esp}
\end{figure}

\begin{figure}
\includegraphics[angle=0, width=0.99\textwidth]{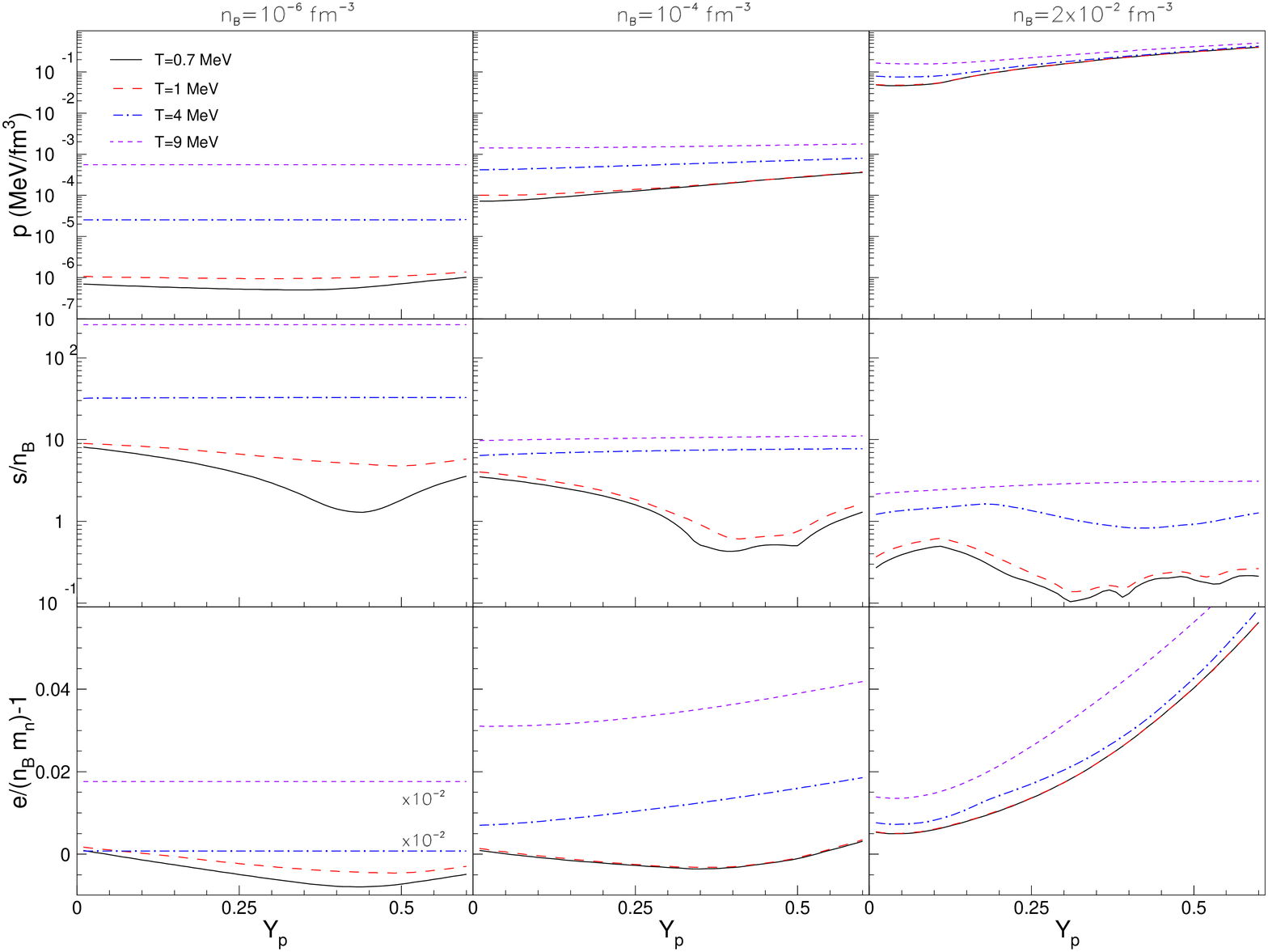}
\caption{Total pressure (top), entropy per baryon (middle) and shifted
energy per baryon (bottom) as a function of proton fraction for fixed values of
the baryonic number density and temperature, see legend.}
\label{fig:total_esp}
\end{figure}

Baryonic and total pressure, entropy and energy per nucleon are shown 
as a function of $Y_p$ in Figs. \ref{fig:baryon_esp} and \ref{fig:total_esp}.
The same thermodynamic conditions as in Figs. \ref{fig:Xnuc} and \ref{fig:NhZh} are considered.
Note that the baryonic pressure already includes the contribution of the Coulomb lattice.
For the sake of convenience the internal energy per nucleon is scaled and shifted 
by the nucleon mass.
The presently considered EoS model  shows features similar to those of other SNA and NSE models
in the literature.
Roughly speaking, two domains may be identified based on the relative 
dominance of homogeneous or clusterized matter.
Whenever matter is mainly composed of free protons and neutrons, 
e.g. low densities and/or high temperatures and/or extreme values of the proton fraction,
the baryonic pressure scales with density and temperature.
When, at the contrary, the thermodynamic conditions are such that an important amount of matter 
is bound in nuclei, the pressure decreases. Even negative values may be reached, due to the Coulomb lattice.
This behavior is easy to understand considering the proportionality between the baryonic pressure and
the total multiplicity per unit volume, see eq. (\ref{eq:pcl}).

The baryonic entropy per nucleon  shows a similar behavior. 
As expected, it increases with increasing temperature and decreasing  $n_B$ as 
more nucleons and light clusters are populated, that is the effective number of degrees of freedom increases.
The most significant decrease of $s_B/n_B$ arises in symmetric matter at low temperatures, where 
matter almost entirely consists of almost isospin symmetric, and thus strongly bound, massive nuclei. 
For low $T$-values, the high sensitivity of isotope population to the global proton fraction, shown 
in Fig. \ref{fig:NhZh}, leads to a wobbling behavior of $s_B(Y_p)$.
At variance with this, dominance of nucleons and light clusters at high temperatures makes $s_B$ 
independent on $Y_p$.

The energy per baryon roughly replicates the $T$- and $n_B$-dependencies of $s_B/n_B$.
This fact is easy to understand considering the  similarity of 
Eqs. (\ref{eq:ecl}) and (\ref{eq:scl}) and can be summarized as follows.
Dominance of heavy clusters, that occurs at low temperature and/or $Y_p \approx 0.5$
and/or high densities, is signaled by small or negative values of $e_B/(n_B m_n)-1$
while dominance of nucleons and light clusters gives high values of the considered quantity. 

Accounting for electron and photon contributions leads to a global increase of the pressure, 
energy and entropy. This is shown in Fig. \ref{fig:total_esp}. We can see that
the highest effects concern the pressure and the internal energy per baryon at high densities.
With the addition of the electron and, to a less extent, photon contributions, 
the total pressure is always positive meaning that a baryonic matter which would be unstable 
if no leptons were present, is stabilized by the presence of the electrons. 
This is the physical reason behind the well-known quenching of the nuclear liquid-gas phase transition 
in stellar matter \cite{RG_PRC_2010}.
The effect on $e_B/(n_B m_n)-1$ is less spectacular and 
trivially due to the increase of electron energy with density.

\begin{figure}
\includegraphics[angle=0, width=0.99\textwidth]{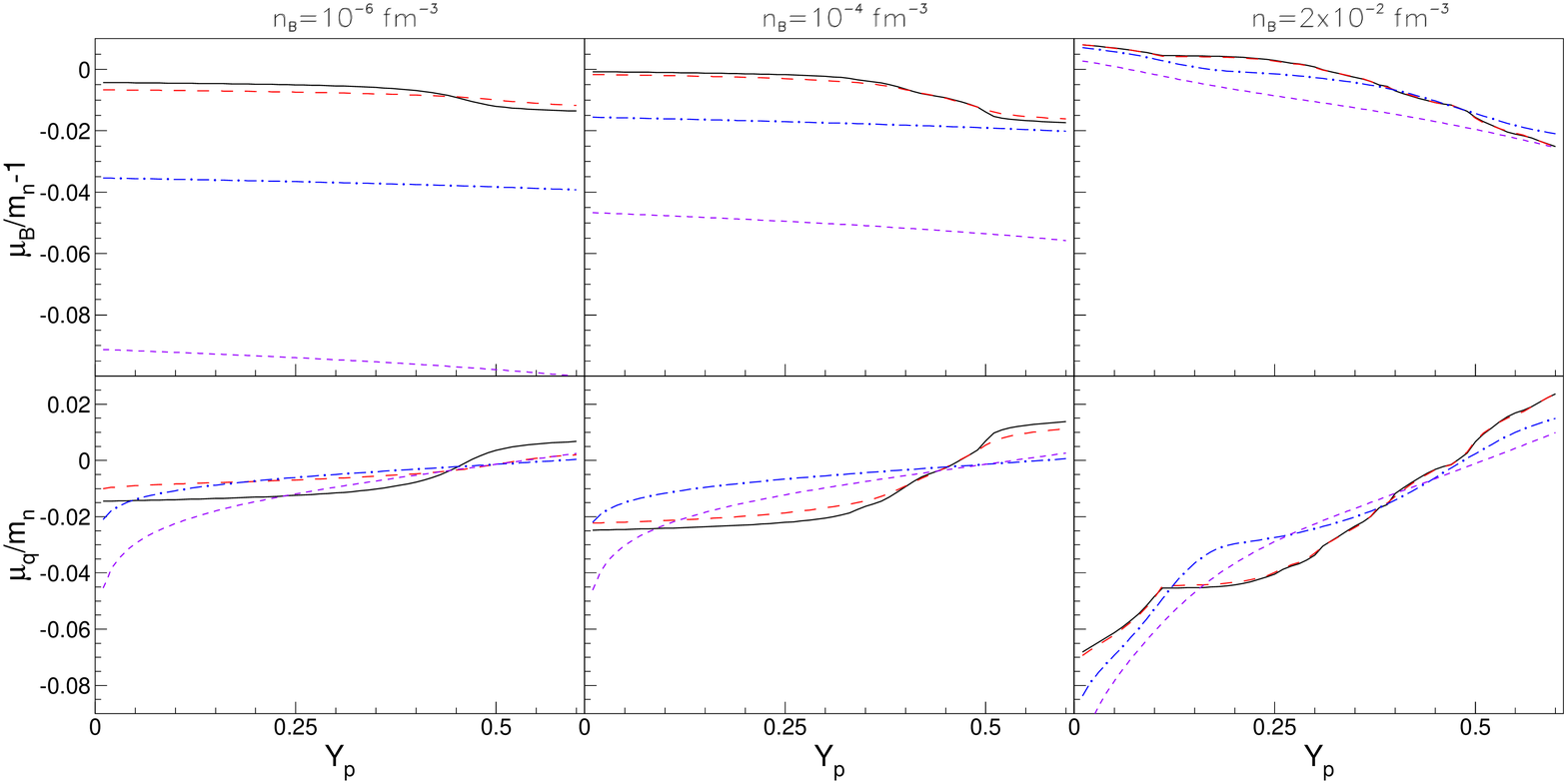}
\caption{Scaled and shifted baryon chemical potential and scaled charge chemical potential
as a function of proton fraction for fixed values 
of the baryonic number density and temperature. 
Legend as in Fig. \ref{fig:baryon_esp}.}
\label{fig:mu}
\end{figure}

Baryon and charge chemical potentials, equal by definition to the neutron chemical potential 
($\mu_B=\mu_n$) and, respectively, the difference between proton and neutron chemical potentials 
($\mu_Q=\mu_p-\mu_n$), are plotted in Fig. \ref{fig:mu} after  subtracting the neutron mass and,
in the case of $\mu_B$, applying a constant shift  for better visibility. 
The same thermodynamic conditions as in Figs. \ref{fig:Xnuc}-\ref{fig:total_esp} are considered.
The monotonic decrease (increase) of $\mu_B$ ($\mu_Q$) with $Y_p$ at constant $n_B$-values
and the evolution of $\mu_B$ with respect to $n_B$ are trivially due 
to the proportionality between one species chemical potential and its density, 
in stable nuclear matter.
The decrease of $\mu_B$ as a function of $T$ recalls the usual behavior of an ideal gas.
On top of these expected behaviors, the heavy clusters component is responsible for non trivial 
small scale variations of $\mu_B$ and $\mu_Q$.

\subsection{Comparison with other NSE models for the EoS}
\label{ssec:comparison}

\subsubsection{General survey}
\label{sssec:general}

As already mentioned,
detailed microphysics input in NSE modelling, e.g. the
nuclear matter energy functional and the cluster definition in terms of maximum baryonic number,
isospin asymmetry and excitation energy, were proven to have little impact on 
thermodynamic quantities such as pressure, energy and entropy densities and
chemical potentials. At variance, a certain sensitivity was identified on the matter composition, 
namely the sharing  between homogeneous and clusterized components, 
as well as the elemental and isotopic abundances.
Two examples in this sense are offered by the inner crust of proto-neutron stars and 
the central element of the core during the late stages of the collapse, when 
$Y_p$-values as low as 0.1 are attained.
In the first case the temperatures are of the order of few MeV and the
baryonic number densities are around $n_0/2-2 n_0/3$.
In the later case, $n_B \lesssim n_0$ and $ T \approx 20$ MeV.
Under these circumstances, other nuclei than those existing 
in the vacuum are expected to be produced, notably extremely neutron rich nuclei, and even 
nuclei beyond drip-lines.
Because of the lack of experimental information on these exotic species, 
the predictions become strongly model dependent.

In order to quantify the model dependence we compare the predictions of the present model, which, in addition to 
AME2012 \cite{Audi_2013}, employs the DZ10 \cite{DZ10} mass model prolonged, for larger isospin asymmetries, 
with a SLy4-based LDM parametrization \cite{Danielewicz}, and introduces a cut-off on the 
internal state density to avoid double counting with the nuclear gas (see \ref{ssec:WScellandcl}),
with those of other two NSE models in literature.
The first considered EoS model is HS DD2 \cite{Hempel_NPA_2010} which, 
in addition to a number of light species (d, t, $^3$He, $^4$He), accounts for 
8979 nuclei ranging from $^{16}$O to $^{339}$136 and extending from the proton drip line to the
neutron drip line. Binding energies are implemented according to the Finite Range Droplet Model 
(FRDM) mass model \cite{FRDM_1997}.
The second EoS model is FYSS \cite{Furusawa_NPA_2017}.
It allows for nuclei far beyond drip lines and nuclei whose proton number may reach values as high as 1000 units.
Binding energies are calculated via a LD-parametrization supplemented by phenomenological 
(temperature-dependent) shell corrections and in-medium modification of the surface energy.
Concerning the unbound nucleon interactions, these models employ RMF functionals,
DD2 \cite{Typel_PRC_2010} (HS DD2) and TM1 \cite{TM1} (FYSS).
Both HS DD2 and FYSS are available on the CompOSE \cite{compose} database.

\begin{figure}
\includegraphics[angle=0, width=0.99\textwidth]{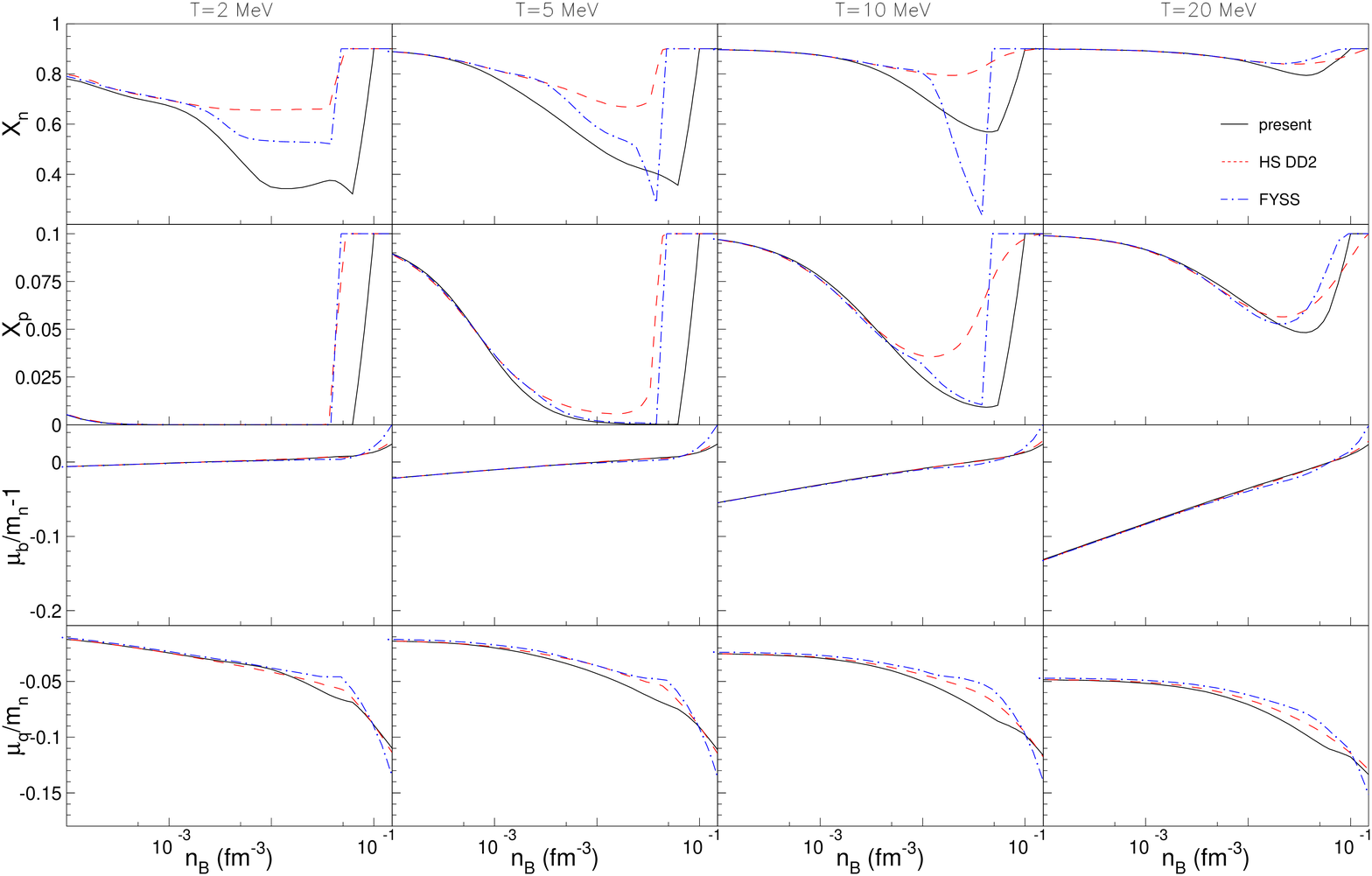}
\caption{(1st and 2nd row panels) Unbound neutrons and protons mass fractions and
(3rd and 4th row panels) scaled baryon and charge chemical potentials as a function
of $n_B$ for $T=2, 5, 10, 20$ MeV and $Y_p=0.1$. The predictions of the present model are
confronted with those of HS DD2 \cite{Hempel_NPA_2010} and FYSS \cite{Furusawa_NPA_2017}.}
\label{fig:comparison_1}
\end{figure}

\begin{figure}
\includegraphics[angle=0, width=0.99\textwidth]{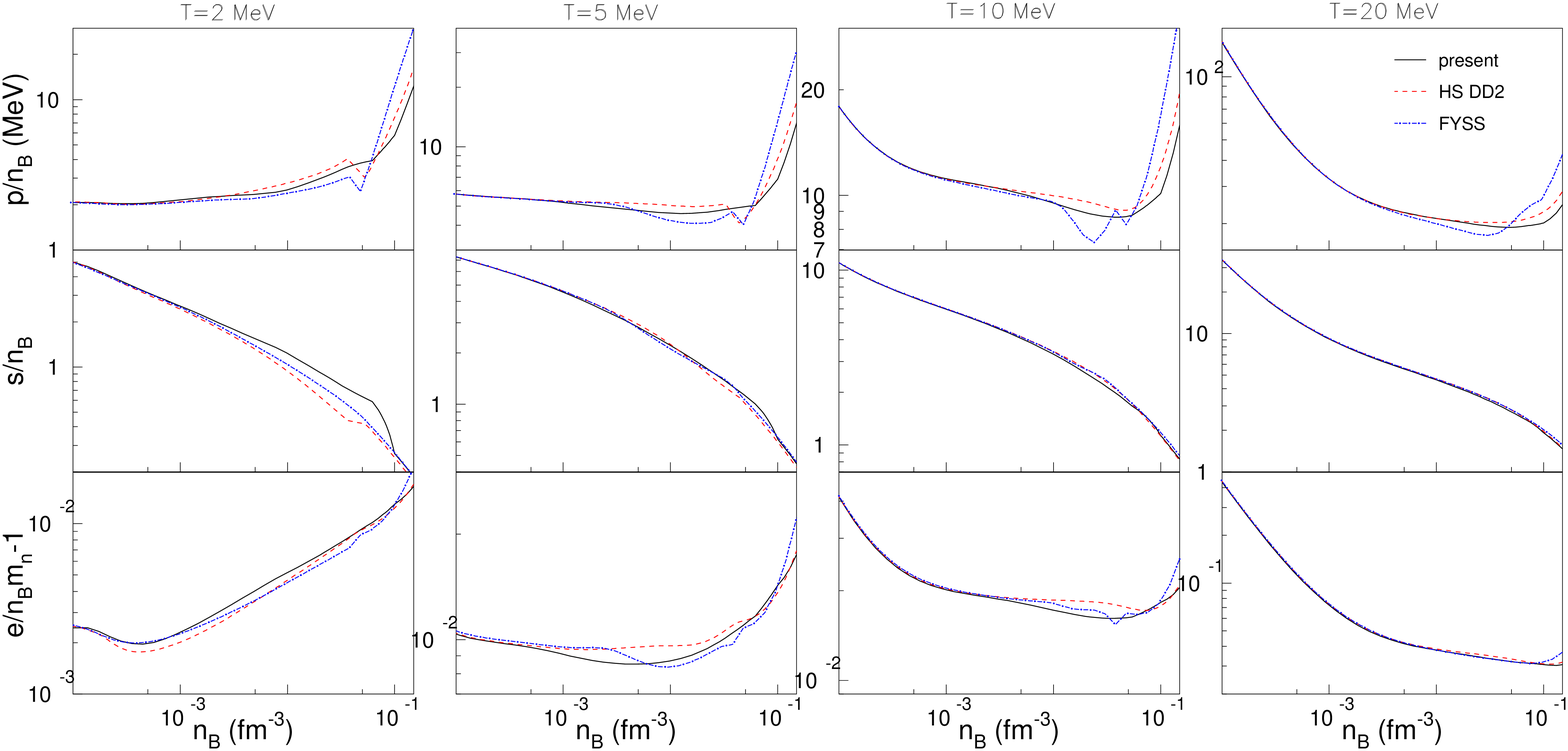}
\caption{The same as in Fig. \ref{fig:comparison_1} for total pressure per nucleon,
entropy per nucleon and energy per nucleon. The last quantity is additionally shifted and scaled
with the neutron mass.}
\label{fig:comparison_2}
\end{figure}

The mass fractions of unbound neutrons and protons, scaled and shifted baryon chemical potential,
scaled charge chemical potential and 
total pressure, energy and entropy per nucleon are  displayed in 
Figs. \ref{fig:comparison_1} and \ref{fig:comparison_2} as a function of the baryonic number density
for $T=2, 5, 10, 20$ MeV and $Y_p=0.1$.
The mass sharing between nuclear clusters is not represented since the information is not available 
for HS DD2 and FYSS.
The neutron mass fraction shows a strong sensitivity to the details of the models, especially
at low temperatures. For instance at $T=2$ MeV and $5 \cdot 10^{-3} ~ {\rm fm^{-3}} \lesssim n_B \lesssim n_t$ 
HS DD2 predicts between 44\% and 86\% and more neutrons than the present model. 
For $T=5$ MeV, the difference amounts to 66\% around $n_B=3 \cdot 10^{-2} ~ {\rm fm}^{-3}$.
The explanation relies in the much limited isospin asymmetry allowed for clusters in HS DD2.
The maximum dispersion among the different NSE models concerning $X_p$ is obtained at $T=10$ MeV
and $1 \cdot 10^{-2} \lesssim n_B \lesssim n_t$, with a maximum ratio $X_p^{HS DD2} /X_p^{present} \approx 4$. 
Concerning the neutrons, the difference would be strongly reduced if we would count the extra neutrons 
beyond the dripline in vacuum as free neutrons. 
Given their small values, the differences in what concerns $X_p$ are not expected to significantly 
influence the electron capture rates and, thus, affect the collapse.
The baryon and charge chemical potential show different features:
$\mu_b$ shows no sensitivity to the details of the models while $\mu_q$  shows a certain sensitivity,
especially at the highest considered densities, where differences up to 50\% are obtained.
Fig. \ref{fig:comparison_2} shows that the other three thermodynamic observables, 
$p/n_B$, $s/n_B$ and $e/n_B m_n -1$, are mostly sensitive to microphysics input at 
$2 \lesssim T \lesssim 10$ MeV and $10^{-3} ~ {\rm fm^{-3}} \leq n_B \leq n_t$.
The much more significant differences that occur at densities higher than the transition density to
homogeneous matter are due to the different EoS and are not relevant for the NSE treatment.

\subsubsection{Detailed chemical composition}

Before the first extended NSE models were developed, less than ten years ago, astrophysical simulations typically
modeled the chemical composition using the Saha equation, which neglects interactions 
among unbound nucleons and between nucleons and nuclear clusters. 
At low densities, this is certainly a very good approximation. 
At high densities excluded volume effects become important but, along a core collapse trajectory, 
their effect tends to be somehow limited by the increasing temperature, which favors the production of
lighter nuclei.

\begin{figure}
\begin{center}
\includegraphics[angle=0, width=0.7\textwidth]{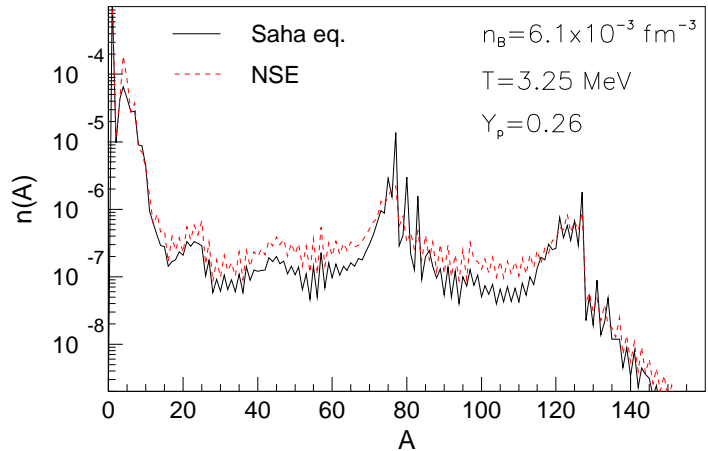}
\end{center}
\caption{Atomic mass number distribution for the last point in the in-fall evolution
of a 25$M_{\odot}$-progenitor considered in Ref. \cite{Juodagalvis_NPA_2010}. The thermodynamic conditions,
mentioned on the figure, correspond to the central element of mass $0.01M_{\odot}$.
Predictions of the present EoS model care confronted with those of the Saha equation.}
\label{fig:nse_saha}
\end{figure}

To quantify the effect of extended NSE, we compare in Fig. \ref{fig:nse_saha} 
the predictions of the present EoS model with those of the Saha equation, 
\begin{equation}
n(A,Z)=g_T(A,Z) \left( \frac{M_{A,Z} T}{2 \pi \hbar^2}\right)^{3/2}
\exp\left[\frac{\left( A-Z\right) \mu_n + Z \mu_p - M(A,Z)+\delta E_{Coulomb}}T    \right],
\end{equation}
where $M(A,Z)$ stands for the nuclear mass 
and the same clusters definition has been employed.
The thermodynamical conditions, $n_B=6.1 \cdot 10^{-3}$ fm$^{-3}$, $Y_p=0.26$ and $T$=3.25 MeV,
correspond to the most compressed state of the central element of mass $0.01M_{\odot}$ of a
25$M_{\odot}$-progenitor, whose in-fall evolution was followed in Ref. \cite{Juodagalvis_NPA_2010}.
The entropy per baryon equals $\approx 1.7$.
The mass distribution predicted by the Saha-equation presents the same features as the NSE-distribution,
including the multi-peak structure induced by shell effects. 
Quantitative differences are nevertheless apparent: 
the most abundant masses have multiplicities that might 
differ by one order of magnitude when one switches from one model to the other and the "valley"
between to peaks gets correspondingly reduced/enhanced, because of conservation laws.

\begin{figure}
\begin{center}
\includegraphics[angle=0, width=0.99\textwidth]{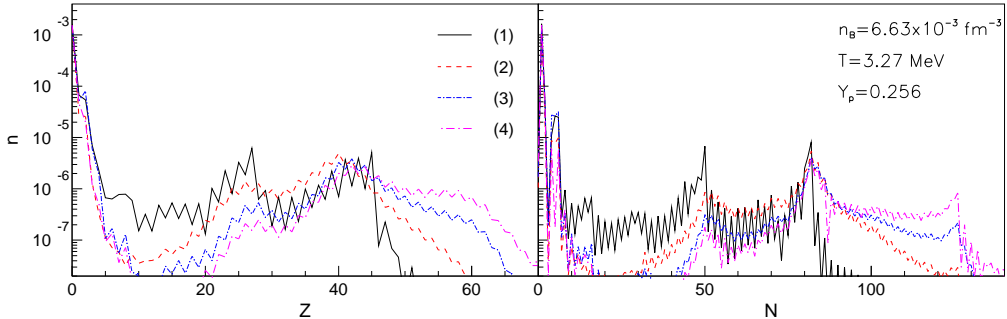}
\end{center}
\caption{Proton (left) and neutron (right) number distributions of nuclei produced under thermodynamic
conditions similar to those of Fig. \ref{fig:nse_saha}.
Predictions of Saha equation (1-3) corresponding to different nuclear cluster definition (see text) are confronted 
with those (4) of Ref. \cite{Hempel_NPA_2010}. 
In all cases experimental data and predictions of FRDM are used for nuclear masses.}
\label{fig:saha_ingredients}
\end{figure}

As already discussed in Ref. \cite{Buyukcizmeci_NPA_2013}, the differences among NSE model predictions
essentially stem from the different fragment definition. 
To get extra insight on the issue we consider here the effect of different assumptions made
on the maximum allowed excitation energy $E^*_{max}$
({\it i.e.} the upper limit of the sum/integral in Eq. (\ref{eq:degen})), and on the level density.
Thermodynamic conditions similar to those in Fig. \ref{fig:nse_saha} are considered.
For nuclear binding energies we use experimental data \cite{Audi_2013} and FRDM \cite{FRDM_1995} predictions
for $2 \leq A \leq 15$ and, respectively, $A \geq 16$.
To keep the modelling as simple as possible, here we restrict ourselves to the predictions of the Saha
equation.
The following cases are investigated:
(1) $E^*_{max}=\min(S_n, S_p)$, as assumed in Ref. \cite{eNSE_PRC2015} and in the present EoS model, 
and $\rho_{A,Z}(E^*)$ of Ref. \cite{Bucurescu2005},
(2) $E^*_{max}=B(A,Z)$, where $B(A,Z)$ is the binding energy as in ref.\cite{Hempel_ApJ_2012}, 
and $\rho_{A,Z}(E^*)$ of Ref. \cite{Bucurescu2005},
(3) $E^*_{max}=B(A,Z)$ and $\rho_{A,Z}(E^*)$ as in Eq. (3) of Ref. \cite{Hempel_NPA_2010}.
The results are plotted in Fig. \ref{fig:saha_ingredients}, together with the predictions (4) of 
the extended NSE (DD2-FRDM) model of Ref. \cite{Hempel_NPA_2010}, as available at \cite{Hempel_Basel}.
Several observations are in order.
Allowing for high excitation energies washes out to a large extent the staggering due to odd-even and shell structure effects,
and also favors heavy nuclei with higher state density.
Different level density formulas lead to different abundancies.
Finally, the curves (3) and (4) compare Saha and extended-NSE.
In all cases, magic nuclei with $Z=28$, $N=50$ and $N=82$ lead to abundance peaks, though their heights 
prove very sensitive to the working hypotheses.

\section{Conclusions}

In this paper we have presented extensive calculations of the composition of supernova matter 
and its thermodynamic properties, in the framework of the extended NSE model of Ref. \cite{eNSE_PRC2015},
whose main formalism is recalled.  
The corresponding EoS database covers large intervals of temperature $0.3 \leq T \leq 50$ MeV, 
baryonic density $10^{-12} \leq n_B \leq 1.5~{\rm fm}^{-3}$, 
and proton fraction $0.01 \leq Y_p \leq 0.6$, with a discretization adapted for 
direct applications in supernova simulation. 
This database is provided in the form of tables following the standards of the CompOSE database 
\cite{compose} (see the Appendix). 
The present NSE-EoS database is complementary to the other, already available, NSE-EoS databases on the 
CompOSE platform.
Indeed it is the first complete NSE table based on a non-relativistic Skyrme energy functional, 
and as such it can be used to quantify the effect of the EoS model dependence on the supernova dynamics. 
A detailed comparison with other available models shows that the most important differences 
among the present modellings concern the treatment of the cluster functionals, 
and this model dependence is more important than the one due to the value of the EoS empirical parameters 
explored for instance in Ref. \cite{Fischer_EPJA_2014}. 
Some of the different treatments can be easily justified. 
This is notably the case of the nuclear pool considered in the probability distribution. 
The same is true for the cluster functional, which must be sophisticated enough to include realistic 
shell effects, which are the essential ingredient that determine the nuclear distributions, 
and therefore the electron capture rates. 
Other differences between the existing models are more subtle and concern the many-body treatment 
inside the WS cell. In particular, the in-medium energy shifts of the light clusters and 
the in-medium modification of the surface tension are important and physically
sound ingredients, which however are difficult to implement consistently in a NSE model. 
As a consequence, different models use phenomenological parametrizations which contain a 
certain degree of arbitrariness. This aspect deserves further progress for a model-independent 
treatment of the microphysics of supernova dynamics.

\section{Appendix: The CompOSE database}

The CompOSE database,
available in the public domain at http://compose.obspm.fr, 
is a repository of EoSs for astrophysics purposes.
Detailed thermodynamic and composition data are provided in standardized 
formats as three dimensional arrays as a function of $T$, $n_B$ and $Y_p$.
Mash points on these quantities are specified in the eos.t, eos.nb, eos.yp files.
The domains covered by the presently discussed EoS model and the number of points
are given in Table \ref{tabel:griddetails}.
\begin{table}
\begin{center} 
\begin{tabular}{cccc}
\hline 
     & $T$       & $n_{B}$ & $Y_p$  \\
\hline 
number of points  & 120 & 140 & 59     \\
minimum           & 0.3 MeV  & 10$^{-12}$ fm$^{-3}$ & 0.01 \\
maximum           & 50.0 MeV  & 1.5 fm$^{-3}$ & 0.60 \\
\hline                                                                                                                                    
\end{tabular}
\caption{Domains of temperature, baryonic number density and charge fraction covered by the present EoS database
and the corresponding numbers of mash points. 
}
\label{tabel:griddetails}
\end{center}
\end{table}

The thermodynamic quantities, stored in eos.thermo, are:
pressure divided by baryon number density $p/n_B$ [MeV],
entropy per baryon $s/n_B$,
scaled and shifted baryon chemical potential $\mu_B /m_n-1$,
scaled charge chemical potential $\mu_Q /m_n$,
scaled electron chemical potential $\mu_{el} /m_n$,
scaled and shifted free energy per baryon $f /(n_B m_n )-1$ and
scaled and shifted energy per baryon $e/(n_B m_n )-1$.
$m_n$ is the nucleon mass, specified in eos.thermo.

The composition data, stored in eos.compo, consists in the 
particle fractions of neutrons ($n_n/n_B$) and protons ($n_p/n_B$)
together with those, $n(A,Z)/n_B$, of the at maximum 500 most probable nuclides whose 
multiplicity per unit volume is not less than $\left(f_{lim} Y_{max} \right)$, where $Y_{max}$ is the  
multiplicity per unit volume of the most abundant nucleus with $A \geq 2$.
For $f_{lim}$, over complementary domains, two values are used: 
$10^{-5}$ and $10^{-8}$.
Note that, because of excluded volume effects, mass and charge conservation equation
are expressed in terms of e-clusters, as already specified in eqs. (\ref{eq:conserv}).


\section*{Acknowledgments}

This work has been partially funded by Pharos, COST Action CA16214.
Ad. R. R. acknowledges useful discussions with Micaela Oertel and kind hospitality from LUTH-Meudon.

\end{document}